\documentclass[aps,pra,twocolumn,superscriptaddress,showpacs,showkeys,amsmath,
amssymb,fleqn,byrevtex]{revtex4}

\usepackage{graphicx}
\usepackage{bm}

\begin{document}

\title{Quantum key distribution for $d$-level systems with generalized Bell states}

\author{Vahid Karimipour}
\email[]{vahid@sina.sharif.edu}
\affiliation{Department of Physics, Sharif University of Technology}

\author{Saber Bagherinezhad}
\email[]{bagherin@ce.sharif.edu}
\affiliation{Department of Computer Science, Sharif University of
Technology\\
P.O. Box 11365-9161, Tehran, Iran}

\author{Alireza Bahraminasab}
\email[]{baramina@physics.sharif.edu}
\affiliation{Department of Physics, Sharif University of Technology}

\date{\today}

\begin{abstract}
Using the generalized Bell states and controlled not gates, we
introduce an enatanglement-based quantum key distribution (QKD) of
$d$-level states (qudits). In case of eavesdropping, Eve's
information gain is zero and a quantum error rate of
$\frac{d-1}{d}$ is introduced in Bob's received qudits, so that
for large $d$, comparison of only a tiny fraction of received
qudits with the sent ones can detect the presence of Eve.
\end{abstract}



\maketitle

\section{Introduction}
As far as classical computation and classical communication are
concerned, binary units of memory and binary logic gates play an
inevitable and natural role due to the inherent simplicity of
Boolean algebra on the one hand and their compatibility with on
and off states of electronic switches on the other hand. With such
classical gates as NOT, AND, and OR, the simplest logical
operations with which we are familiar in everyday life, and also
any binary function can be implemented. They are also quite simple
to design electronically. However in quantum computation and
communication (see \cite{nielson, preskill} and references
therein), the main resources that have the potential of surpassing
our conventional classical methods, are quantum parallelism (for
massive computation), non-locality and entanglement (for
communication) and uncertainty relations (e.g. for Quantum Key
Distribution among other things). For utilizing these resources,
two level quantum states are by no means inevitable. Only
considerations of quantum hardware should decide between using
2-level or multi level states. At present a major difficulty in
quantum computation is the limit on the number of qubits that can
be coupled experimentally \cite{steane}. Although it may be easier
to construct universal gates for qubits than for qudits, the use
of d-dimensional systems or qudits has the advantage that compared
to qubits, fewer systems should be coupled to obtain a given
dimensionality of the Hilbert space. Apart from practical
considerations, it will enhance and deepen our understanding of
the subject if we try to re-formulate quantum computation and
communications in a dimension-free context. In view of this,
various authors have tried to generalize some of the algorithms,
protocols, or error correcting codes of two level quantum
computation to arbitrary dimensional Hilbert spaces
\cite{gottesman, albers1, albers2, albers3, bartlett, knill, chau,
hb1, hb2, cerf}. Consequently one sees in the literature that the
same basic tool (i.e. a generalized gate) has been
defined independently in several works.

For example the generalization of one of the basic gates of
quantum computation, thats the controlled not gate appears to have
been given independently by a number of authors under different
names \cite{gottesman, albers1, albers2, bartlett, mermin, kbb2}.
The same is also true for the generalized Hadamard gate. Specially
in \cite{albers1} an experimental realization of the generalized
XOR gate to $d$-levels has been proposed, where the number $n$ of
photons in an electromagnetic mode signifies the state $|n\rangle
,\ n = 0,\ \ldots,\ d-1$ and a Kerr interaction between these photons
and their fourier transform is used to induce the generalized XOR
gate on the states.

In this paper we are concerned with a protocol of Quantum Key
Distribution (QKD) and its generalization to states of arbitrary
dimension. Quantum Cryptography (QC) which is based on very simple
ideas and yet not far from real applications as the other
highlights of quantum computations are (like factoring large
integers) is one of the most promising
areas of research in Quantum Computation and Information.

Particularly interesting is that in QC one tries to turn the
apparently negative or counter-intuitive rules of quantum
mechanics, which has resulted in epistemological debates in the
past decades, into absolutely useful devices for engineering
applications. One such concept has been the uncertainty principle,
or the fact that observation or measurement perturbs the
observable. This rule has been utilized in a most beautiful
application in the form of BB84 protocol for QKD \cite{bb84},
where bits of a Key prepared by two legitimate parties, in the
form of spin or polarization of particles in random bases, are
inevitably perturbed by a non-legitimate third party. (For a
review on QC including many theoretical and
practical issues, see \cite {rev}.)

Another nonclassical and counter-intuitive concept, has been the
concept of non-locality and entanglement which has found even
wider applications, to the extent that nowadays a major problem
about non-locality is not how to interpret it, but how to measure
it like other useful resources as energy and momentum.

The first entanglement-based protocol of QKD has been the work of
Eckert \cite{ekert}, which later was shown to be equivalent to the
original BB84 protocol \cite{bb84'},( see \cite{rev} for finer
details). Two other QKD protocols which have used entanglement in
an essential way has been reported in \cite{cabello} and
\cite{zlg}. The first of these uses entanglement swapping via Bell
measurements to securely transfer a key and has been generalized
to $d$-level systems in \cite{kbb2}. The second is based on local
gate operations on a reusable EPR pair, where Alice tries to hide
the secure data by entangling each bit with the EPR pair, sending
the bit to Bob who can disentangle the bit and read the data. The
strategy of Eve is to somehow entangle herself with the whole
state of the EPR and the bit by suitable operations and
find access to the data, without being revealed by Alice and Bob.

The aim of this paper is to generalize this second protocol to
higher dimensional states and at the same time give a clear
exposition of its basics.

For the sake of brevity, we will not
go into the details of 2-level protocol. For this, the reader can
either consult reference \cite{zlg}, or else go through
the next sections and at each step set $ d = 2 $.

The basic advantage of this protocol is that as we will show, not
only Eve's presence will be detected by Alice and Bob, but also
her information gain is zero, compared to the 50 percent
information gain in the BB84 protocol. This is true in every dimension, but as we will show
Eves presence introduces a higher Quantum Bit Error Rate, in higher dimensional states, so that
her presence can be detected more easily.

The structure of this paper is as follows. In section 2 we first
review some known and new facts about the generalized Bell states,
and the generalizations of CNOT and the Hadamard gates to qudits.
In section 3 which has been divided into several subsections, we
generalize the QKD scheme of \cite{zlg} to $d$-level systems, and
discuss the security of the protocol against some individual
attacks. We show that the information gain of Eve is actually zero
and show how the intervention of Eve introduces an error rate of
$\frac{d-1}{d}$ into the data received by Bob, and greatly
enhances the chance of her detections by the legitimate parties.
In all this we are concerned only with theoretical considerations
and do not consider practical issues, or any rigor in proving
security, all these are important but should be considered in
separate works. Finally in section 4 which concludes the paper, we
discuss a possible route for generalization of our results to the
continuous variables. Some of the calculations which are not
detailed in the main text, are collected in the appendix.

\section{States, and gates for $d$-level systems}
For qudits, a generalization of the familiar Bell states, has been
introduced in \cite{bell1, bell2, bell3, albers1}. These are a set of $d^2$
maximally entangled states which form an orthonormal basis for the
space of two qudits. Their explicit forms are:

\begin{equation}\label{bells}
|\Psi_{m,n}\rangle:=\frac{1}{\sqrt{d}} \sum_{j=0}^{d-1}
\zeta^{nj}|j,j+m\rangle
\end{equation}
where $ \zeta = e^{\frac{2\pi i}{d}}$ and $m $ and $ n $ run from
$ 0 $ to $ d-1$. These states have the properties
$\langle\Psi_{m,n}| \Psi_{m',n'}\rangle =
\delta_{n,n'}\delta_{m,m'}$ (orthonormality) and $\text{trace}_2(
|\Psi_{m,n}\rangle\langle\Psi_{m,n}|) = \frac{1}{d} \openone$ (maximal
entanglement). The following operators \cite{bell1, bell2, bell3, gottesman} are
also useful, since they play the analogous role of Pauli operators
for qudits:
\begin{equation}\label{pauli}
U_{m,n} = \sum_{j=0}^{d-1} \zeta^{nj}|j+m\rangle\langle j|
\end{equation}
For example, given the entangled state $ |\Psi_{0,0}\rangle$, only
one of the parties, say Alice, can generate any Bell state $
|\Psi_{m,n}\rangle $ by acting on $ |\Psi_{0,0}\rangle $ with $
U_{m,n}$, i.e:

\begin{equation}\label{paulionbell}
(\openone\otimes U_{m,n})|\Psi_{0,0}\rangle = |\Psi_{m,n}\rangle
\end{equation}
One should however note that contrary to the Pauli operators, the
operators $ U_{m,n}$ are not necessarily Hermitian.

One can also
generalize the Hadamard gate which turns out to be quite useful in
manipulating qudits for various applications \cite{gottesman,
 albers1, mermin}.
\begin{equation}\label{hadamardket}
  H :=\frac{1}{\sqrt{d}} \sum_{i,j=0}^{d-1} \zeta^{ij}|i\rangle\langle j|
\end{equation}
where $ \zeta = e^{\frac{2\pi i}{d}}$. This operator is really not
new and it is known as the quantum Fourier transform when $d = 2
^n$. In that case it acts on $ n $ qubits. Here we are assuming it
to be a basic gate on one single qudit, in the same way that the
ordinary Hadamard gate is a basic gate on one qubit. This operator
is symmetric and unitary ($ H H^* = \openone $), but not hermitian.

To
generalize the NOT and the CNOT gates, we note that in the context
of qudits, the NOT gate, is basically a mod-2 adder. For qudits
this operator gives way to a mod-$d$ adder, or a Right-Shift
gate,\cite{gottesman, albers1, bartlett, mermin, kbb2}
\begin{eqnarray}\label{rightandleftshift}
  R|j\rangle = |j+1\rangle\text{mod} & d \\
  R^{-1}|j\rangle \equiv L|j\rangle = |j-1\rangle\text{mod} & d ,
\end{eqnarray}
where $ L $ has been used to denote a left shift. Note that $ R^d
= \openone$, compared to $ NOT^2 = \openone$.

For every unitary operator $ U $ the controlled gate $ U_c $ which
acts on the second qudit conditioned on the first qudit is
naturally defined as follows:
\begin{equation}\label{controlledgate}
  U_{c}(|i\rangle\otimes|j\rangle) = |i\rangle\otimes U^i|j\rangle
\end{equation}
Note the difference with the qubit case. In the qubit case a
controlled operator acts only if the value of the first bit is 1,
here it acts $ i $ times if the value of the first qudit is $ i $.
(Sometimes it is said that a controlled operator is like an \textit{
if statement} in classical computation \cite{nielson}. If we take
this statement literally, then a controlled operation for $d$-level
states acts like a loop.)  In particular the controlled shift
gates which play the role of CNOT gate, act as follows:
\begin{equation}\label{controlledshift}
  R_c |i,j\rangle = |i,j+i\rangle \hskip 1cm L_c |i,j\rangle = |i,j-i\rangle
\end{equation}
Every function $ f $ from $\{0,1, \ldots, d-1\}^n \rightarrow
\{0,1, \ldots, d-1\}^m $ is made reversible by the definition
$f_r(\mathbf{x},\mathbf{y}) = (\mathbf{x}, f(\mathbf{x})+\mathbf{y}) $ where all
additions are performed mod d. In quantum circuits such a function
is implemented by a unitary operator $ U_f|\mathbf{x},\mathbf{y}\rangle
:= |\mathbf{x}, f(\mathbf{x})+\mathbf{y}\rangle$ where $\mathbf{x} \in \{0,
1, \ldots, d-1\}^n$ and $\mathbf{y} \in \{0, 1, \ldots, d-1\}^m$.
Note that here and in all that follows addition of multi dits is
performed dit-wise and mod $d$.

Quite analogously to the q-bits, the Hadamard and the Controlled
Shift gates can generate all the Bell states $\{|\Psi_{m,n}\rangle
\}$ from the computational basis states $\{|m,n\rangle
\}$ \cite{albers1}:

\begin{equation}\label{generationofbell}
R_c(H\otimes\openone)|n,m\rangle = |\Psi_{m,n}\rangle
\end{equation}

\begin{figure}
\includegraphics{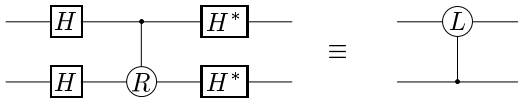}
\caption{\label{fig:HHgates} Circuit identity for $d$-level gates}
\end{figure}

Many other properties of these gates are simply carried over from
the case of q-bits to the general case with appropriate
modifications. For example one can check the validity of the
circuit identity in fig. (\ref{fig:HHgates}).

\section{An entanglement-based protocol of QKD for $d$-level states}
In this section we generalize an entanglement-based protocol of
quantum Key Distribution first put forward in \cite{zlg} to
$d$-level states and perform further analysis of the method.
\subsection{QKD in the absence of Eve}
The starting point of this protocol is the sharing of a Bell state
$ |\Psi_{00}\rangle = \frac{1}{\sqrt{d}}\sum_{j=0}^{d-1}
|j,j\rangle_{a,b} $ by Alice and Bob. The qudit to be sent is
denoted by $q$, which is encoded as a basis state $|q\rangle_{k}$.
Throughout the paper we use the subscripts $a, b, k$ and $e$ for
Alice, Bob, key and Eve respectively. The basic idea, neglecting
considerations of Eve's attack for now, is that Alice performs a
controlled-right shift on $|q\rangle_{k}$ and thus entangles this
qudit to the the previously shared Bell state, producing the state
\begin{equation}
|\Phi\rangle = \frac{1}{\sqrt{d}} \sum_{j=0}^{d-1}
|j,j,q+j\rangle_{a,b,k}.
\end{equation}
She then sends the qudit to Bob. By this operation she is hiding
the qudit $q$ in a completely mixed state, since $ \rho_{k} :=
\text{tr}_{a,b}|\Phi\rangle|\langle \Phi| = \frac{1}{d}\openone_{k}$. At the
destination, Bob performs a controlled-left shift on the qudit and
disentangles it from the
Bell state, hence revealing the value of $q$ with certainty.

Note that in contrast with the BB84 protocol, here the key is not
determined a posteriori and randomly, hence a larger transfer rate
is possible.
\subsection{An individual attack by Eve}
\begin{figure*}
\includegraphics{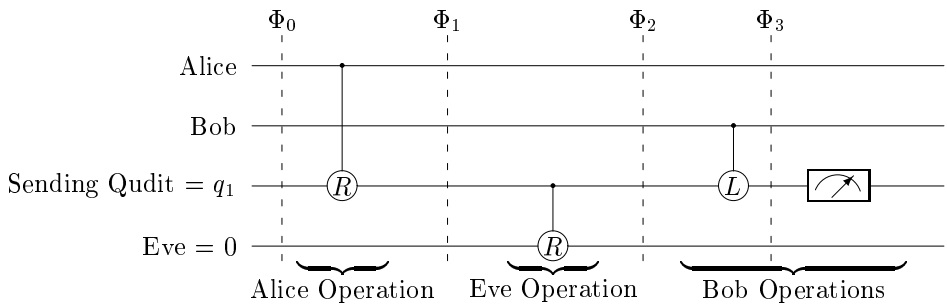}
\caption{\label{fig:eve1} Eve attack for the first qudit}
\end{figure*}

A possible conceivable attack by Eve(e) is that she entangles her
state to those of Alice, Bob and the intercepted key so that after
Bob's measurement of the qudit, she can obtain partial information
about the qudit. The best way to describe and visualize the
protocol is to refer to fig. (\ref{fig:eve1}), where the qudits
are drawn as lines and states at each stage are shown
explicitly.

The strategy that Eve follows should be described separately for
the first qudit and the rest of the qudits. For the first qudit,
she performs no measurement and proceeds so that her qudit gets
entangled with the Bell state of Alice and Bob at the end of the
process. For this she uses a controlled  right-shift on her qudit
conditioned on the value of the first qudit being sent (see fig.
(\ref{fig:eve1})). The states at various stages are as follows,
where in each ket the qudits refer respectively from left to right
to Alice(a), Bob(b), the key(k) and Eve(e):
\begin{eqnarray}\label{evefirstbit}
|\Phi_0\rangle &=& \frac{1}{\sqrt{d}} \sum_{j=0}^{d-1} |j,j,q_1,0\rangle_{a,b,k,e} \\
|\Phi_1\rangle &=& \frac{1}{\sqrt{d}} \sum_{j=0}^{d-1} |j,j,q_1+j,0\rangle_{a,b,k,e} \\
|\Phi_2\rangle &=& \frac{1}{\sqrt{d}} \sum_{j=0}^{d-1}
|j,j,q_1+j,q_1+j\rangle_{a,b,k,e}
\end{eqnarray}
Note that choice of $|0\rangle $ for Eve's original state is quite
arbitrary. Her strategy works with any other choice. In the last
stage when Bob performs his Left-Shift gate, he produces the state
\begin{equation}
|\Phi_3\rangle = \frac{1}{\sqrt{d}} \sum_{j=0}^{d-1}
|j,j,q_1,q_1+j\rangle_{a,b,k,e}
\end{equation}
and thus disentangles the key and measures correctly the value of
its first dit $q_1$. However his shared Bell state with Alice has
now been left entangled with the state of Eve, which is used again
by Alice and Bob (unaware of the entanglement with Eve) for the
next round (i.e; for sending the dit $q_2$ of the key). Thus for
the next round the state that Alice and Bob will start with is:
\begin{equation}\label{abestate}
|\Psi_0\rangle = \frac{1}{\sqrt{d}} \sum_{j=0}^{d-1}
|j,j,q_2,q_1+j\rangle_{a,b,k,e}
\end{equation}

Note that we are assuming that Alice and Bob do not have access to
a reservoir of Bell states, the later being supposedly expensive.
Thus they are using one Bell state for sending the whole key or at
least a considerable fraction of it.

\begin{figure*}
\includegraphics{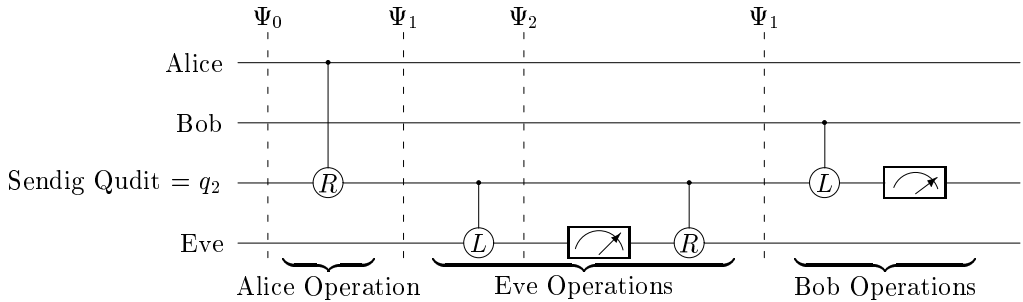}
\caption{\label{fig:eve2} Eve attack for next qudits}
\end{figure*}

It is important to note that Eve modifies her strategy for the
next dits, by first performing a left-shift, measuring her qudit
and then performing a right-shift on her qudit. The rest of the
process is like that for the first qudit (see fig.
(\ref{fig:eve2})). The various states in different stages shown in
the figure are as follows:
\begin{eqnarray}\label{evesecondbit}
|\Psi_0\rangle &=& \frac{1}{\sqrt{d}} \sum_{j=0}^{d-1} |j,j,q_2,j+q_1\rangle_{a,b,k,e} \\
|\Psi_1\rangle &=& \frac{1}{\sqrt{d}} \sum_{j=0}^{d-1} |j,j,q_2+j,j+q_1\rangle_{a,b,k,e} \\
|\Psi_2\rangle &=& \frac{1}{\sqrt{d}} \sum_{j=0}^{d-1}
|j,j,q_2+j,q_1-q_2\rangle_{a,b,k,e}
\end{eqnarray}
At this stage Eve who has disentangled her qudit from the rest of
the state measures her own qudit to be $ q_1 - q_2 $. She then
performs the controlled right shift on her qudit to restore the
original state $|\Psi_1\rangle$. At the destination Bob again
proceeds as before, performs his left shift and measures the value
of $q_2$, leaving the state of Alice, Eve and his own state, in an
entangled
state ready for use in the next round.

In this way Eve intercepts the qudits $$q_1 - q_2, q_1 - q_3, 
q_1 - q_4, \ldots$$  from which she can infer all the sequence by
checking $ d $ possible values
for $q_1$.

Note that for each qudit, Eve is effectively doing an
intercept-recent strategy, however she does not intercept the
value of the qudit (say $q_2$) sent by Alice, but she measures a
value $q_2 - q_1$, where $q_1$ is the value of the first sent
qudit which has been intercepted in an earlier stage.

\subsection{Protection against Eve's intervention}
To protect this protocol against this kind of attack, Alice and
Bob proceed as follows: Before sending \textit{each} of the qudits,
Alice and Bob act on their shared Bell state by the Hadamard gates
$ H $ and $H^*$, respectively. The key point is that a Bell state
$|\psi_{0,0}\rangle$ disentangled from the outside world is
unchanged under this operation, while a state entangled with
outside is not:
\begin{equation}\label{HH}
  (H\otimes H^*)|\Psi_{0,0}\rangle = |\Psi_{0,0}\rangle.
\end{equation}
In the absence of intervention of Eve, this extra operation has no
effect on the protocol.

In fact the shared Bell state is unchanged under more general
operators of the form $ U\otimes U^*$, where $U$ is any unitary
operator. We will investigate this possibility in appendix B.

It is clear from fig. (\ref{fig:eve1}), that for the first qudit
nothing changes. However for the second qudit and other qudits,
essential changes occur in the intermediate states in the process.
As we will see, in this way Alice and Bob can prevent Eve from
getting any useful information. The entangled state of Alice, Bob
and Eve remained from the first round is
\begin{equation}\label{ee0}
|\chi\rangle = \frac{1}{\sqrt{d}} \sum_{j=0}^{d-1}
|j,j,j+q_1\rangle_{a,b,e}
\end{equation}
When Alice and Bob perform their Hadamard gates on their qudits,
this state changes to
\begin{equation}\label{ee1}
|\widetilde{\chi}\rangle = \frac{1}{\sqrt{d}} \sum_{i,j,k=0}^{d-1}
H_{i,j}H_{k,j}^*|i,k,j+q_1\rangle_{a,b,e}
\end{equation}
Thus the second round of the protocol after Alice inserts the
second dit of the key, starts with the state
\begin{equation}\label{ff}
|\widehat{\Psi_0}\rangle = \frac{1}{\sqrt{d}} \sum_{i,j,k=0}^{d-1}
H_{i,j}H_{k,j}^*|i,k,q_2,j+q_1\rangle_{a,b,k,e}
\end{equation}
The state $ |\widehat{\Psi_1} \rangle $ which results after
Alice's controlled $R$ operation will be:
\begin{equation}\label{HHevesecondbit}
|\widehat{\Psi_1}\rangle = \frac{1}{\sqrt{d}} \sum_{i,j,k=0}^{d-1}
H_{i,j}H_{k,j}^*|i,k,i+q_2,j+q_1\rangle_{a,b,k,e}
\end{equation}
The qudit is now sent to Bob. We will show in appendix A, that
this new state has no information for Eve. In fact we will show
that, the density matrix of her system and the qudit will be
\begin{equation}\label{density}
\rho_{k,e} = \frac{1}{d^2} \openone_{k}\otimes\openone_{e}
\end{equation}
Therefore under any unitary operation on her qudit and the sent
qudit whether it be the controlled $L$ gate used for the first
round or a more complex cleverly chosen operator, she will not be
able to get useful information from the intercepted data. More
generally it is hardly possible for Eve that by a quantum
operation derived from suitable interactions with her ancillas,
can derive any useful information from this density matrix.
\subsection{The information gain of Eve}
The above situation is analogous to the case of BB84 protocol,
where with respect to any basis chosen by Eve, the density matrix
of the qubits intercepted by Eve are identity matrices. However
there is one major difference in that in the BB84 protocol and its
variations and generalizations to higher dimensional states, the
protocol ends up with a public announcement of the bases of Alice
and Bob, from which Eve finds that she has intercepted a fraction
of the qubits or qudits correctly. Therefore Eve finds partial
information about the key and only her revealing by Alice and Bob
saves those protocols. Here we will show that the information gain
of Eve is actually zero and she obtains no information at all
about the key. The mean information gain per bit of Eve $I$, is
the difference between two relative entropies and is interpreted
as the percentage of bits which are saved when Eve wants to write
the data of Alice from her own intercepted data \cite{rev}. We
have:
\begin{equation}\label{information}
  I = H_{a\text{ priori}} - H_{a\text{ posterieri}}
\end{equation}
Assuming that Alice sends the qudits uniformly, we have: $ H_{a
\text{priori}}= \log_2 d $. We also have:
\begin{equation}\label{posterieri}
  H_{a\text{ posterieri}} = - \sum_{r} p(r)p(i|r) \log_2 p(i|r)
\end{equation}
where $p(r)$ is the probability that Eve receives a dit value of
$r$ and $p(i|r)$ is the a posterieri probability that Alice has
sent a dit value of $i$ given that Eve has received a dit value of
$ r$. The later can be easily calculated from Bayes's formula
\begin{equation}\label{bayes}
  p(i|r) = \frac{p(i)p(r|i)}{\sum p(i)p(r|i)}
\end{equation}
Since Alice is assumed to send the dits uniformly, we have $p(i) =
\frac{1}{d}$ and since the density matrix of Eve is unity, $p(r|i)
= \frac{1}{d}$, thus we find: $p(i|r) = \frac{1}{d}$. Inserting
all this in (\ref{posterieri}) we find that $H_{a\text{ posterieri}} =
\log_2 d = H_{a\text{ priori}}$ and hence zero information gain for
Eve. This is a very interesting property of this protocol compared
with the BB84 protocol and its variations or generalizations to
higher dimensional systems where the information gain of the Eve
is non-zero. In fact in the BB84 protocol the 50 percent
information gain is due to those occasions where the basis of Eve
happens to be the same as the publicly announced bases of Alice
and Bob. Here there is no public announcement of any kind and so
all the dits that Eve measures
are really worthless at the end of the protocol.

\subsection{Detection of Eve}
At this stage we want to show how Alice and Bob can infer the
presence of Eve from comparison of their data. Although by no
unitary operation on her system and the key, she can gain
information from the key, she may want to use a clever operation
to reduce as much as possible the Quantum Bit Error Rate (QBER)
introduced into the data received by Bob, and hence her chance of
being detected. The QBER depends on her choice of the operation.
Suppose that she performs the same sequence of (Controlled $R$ +
measurement + controlled $L$ ) operations that she was doing in
her successful attacks. It is straightforward to see that with the
preservation of Hadamard gates, the new state that reaches Bob,
provided that the qudits $q_1 $ and $q_2$ have been sent and Eve
has measured a qudit value of $q$ in the second round is:
\begin{eqnarray}\label{afterevemeasure}
|\Phi_3\rangle =
\sum_{i,k=0}^{d-1}&&H_{i,i+q_2-q_1+q}H^*_{k,i+q_2-q_1+q}\nonumber\\
&&|i,k,i+q_2,q\rangle_{a,b,k,e}
\end{eqnarray}
After Bob performs his controlled-$L$ operation, the final state
ready for measurement will be:
\begin{eqnarray}\label{final}
|\Phi_4\rangle =
\sum_{i,k=0}^{d-1}&&H_{i,i+q_2-q_1+q}H^*_{k,i+q_2-q_1+q}\nonumber\\
&&|i,k,i+q_2-k,i+q+q_2\rangle
\end{eqnarray}
It is again a simple computation to find the density matrix
corresponding to the Key space from this state, which turns out to
be
\begin{equation}\label{final2}
\rho_{k} := \text{tr}_{a,b,e}|\Phi_4\rangle \langle \Phi_4| =
\frac{1}{d}\openone_{k}.
\end{equation}
This means that Bob measures all the values of the key qudit with
equal probability and his chance of getting the correct qudit is
$\frac{1}{d}$. Hence the QBER introduced into the data by Eve's
intervention is $\frac{d-1}{d}$.

\section{Discussions}
We have studied a protocol of quantum key distribution for
$d$-level systems based on shared entanglement of a reusable Bell
state and have shown that in this protocol, the information gain
of Eve is zero and the Quantum Bit Error Rate (QBER) introduced by
her interception into the data received by Bob is $\frac{d-1}{d}$.
The situation is similar to the generalizations of the BB84
protocol to higher dimensional states \cite{hb1,hb2,cerf} in which
the larger the number of states, the larger is the QBER, which in
turn may be larger than any noise already present in the channel.
This later fact seems to be an advantage in terms of the security
of the key distribution scheme \cite{Ho1}. These results are based
only on the analysis of a direct individual attack by Eve. It may
be interesting to study further types of attacks and to establish
theoretical bounds to the information gain and the QBER in this
protocol or go through a general analysis along the lines that
have been followed for the BB84 protocol in \cite{Ho1, Ho2, Ho3}
and to see if this protocol has an unconditional security or
not.

Another route for extending our results is to consider the
continuous variables. There has been a lot of interest toward
quantum computation and quantum communication with continuous
variables in the past couple of years (see \cite{cont1, cont2} and
references therein), where instead of bits, information may be
stored in infinite dimensional states like position or momentum of
a particle or amplitude of an electromagnetic field. Part of this
interest derives from the fact that it has been shown that a
combination of optical devices like phase shifters and beam
splitters may be sufficient to act as a set of universal gates.
Therefore many algorithms and protocols have been re-studied for
continuous variables \cite{cont2}. Now that we have a QKD protocol
for $d$-level states for any $d$, a natural question arises whether
it is possible to go to a proper continuous limit and define the
above process for continuous variables. We can simply replace the
discrete states $|j\rangle $ with continuous variables $|x\rangle
, -\infty < x < \infty $ and $\zeta = e^{\frac{2\pi i}{d}} $ with
$ \zeta = e^{2 \pi i}$ in all the formulas for states and
operators to adapt the protocol to the continuous variables. In
all stages we need also change summations to integrations:
\begin{equation}\label{replace}
  \frac{1}{\sqrt{d}}\sum_{0}^{d-1}\ \ \ \rightarrow \ \ \
  \frac{1}{\sqrt{2\pi}}\int_{-\infty}^{+\infty} dx
\end{equation}
Following these we will find  the generalized Bell states in the
continuous case:
\begin{equation}\label{cont1}
|\Psi_{\alpha, \beta}\rangle = \frac{1}{\sqrt{2\pi}}\int e^{i\beta
x}|x,x+\alpha\rangle dx
\end{equation}
where $\alpha $ and $\beta$ are continuous labels ranging from $
-\infty $ to $+ \infty$ and $ |x\rangle$ is a continuous state
like position and all the integrals now and hereafter are over the
real line. These states are normalized in the sense that
\begin{equation}\label{cont2}
\langle\Psi_{\alpha, \beta}|\Psi_{\alpha', \beta'}\rangle =
\delta(\alpha - \alpha') \delta(\beta - \beta')
\end{equation}
and are maximally entangled in the sense that $$
\text{trace}_2(|\Psi_{\alpha, \beta}\rangle\langle \Psi_{\alpha, \beta}|)
=  \frac{1}{2\pi}\int_{-\infty}^{+\infty} |x\rangle\langle x| dx
.$$ The generalization of the Hadamard operator is nothing but the
Fourier transform operator which has already been used in
\cite{cont2} to generalize the Grover algorithm \cite{grover} to
continuous domain.
\begin{equation}\label{cont3}
H |x\rangle = \frac{1}{\sqrt{2\pi}}\int e^{i x y} |y\rangle dy
\end{equation}
The controlled right shift operator now takes the form
\begin{equation}\label{cont5}
R_c |x, y\rangle = |x, x+y\rangle
\end{equation}
which as an operator takes the particularly simple form
\begin{equation}\label{cont4}
R_c = e^{-i X \otimes P}
\end{equation}
This operator has also appeared already in  \cite{albers1}. To
define the form of the protocol for the continuous variables, it
is enough to modify all the states in various stages of the
protocol as stated above. It may then be practically more feasible
to really implement this protocol by optical means.


\appendix

\section{}\label{apx:a}
In this appendix we show that Eve can not counteract
the action of the Hadamard gates by replacing her controlled shift
gate by any other unitary operator or even by any quantum
operation. Therefore any measurement of her system or the
intercepted qudit will reveal nothing to her.

When Eve intercepts the sent qudit, she will have access to the
last two parts of the following state:
\begin{equation}\label{appendix1}
|\Psi_2\rangle = \frac{1}{\sqrt{d}} \sum_{i,j,k = 0}^{d-1}
H_{i,j}H_{k,j}^*|i,k,i+q_2, j+q_1\rangle_{a,b,k,e}
\end{equation}
One can now find the density matrix of Eve and the sent qudit from
$ \rho_{k, e} = \text{tr}_{a,b}(|\Psi_2\rangle \langle \Psi_2|)$. Using
the primed dummy indices like $ i',j'...$ for the bra state
$\langle \Psi_2|$ we have:
\begin{eqnarray}
\rho_{k, e} = \frac{1}{d}\sum&& H_{ij}H^*_{kj}
H^*_{i'j'}H_{k'j'}\delta_{ii'}\delta_{kk'}\nonumber\\
&&|i+q_2, j+q_1\rangle\langle i'+q_2,j'+q_1|
\end{eqnarray}
Summing over $ i', k'$ we find:
\begin{eqnarray}
\rho_{k, e} = \frac{1}{d}\sum&&H_{ij}H^*_{kj}
H^*_{ij'}H_{kj'}\nonumber\\
&&|i+q_2,j+q_1\rangle\langle i+q_2,j'+q_1|
\end{eqnarray}
Summing over $k$ and using the symmetry and unitarity of $H$
($\sum H^*_{kj}H_{kj'}= \delta_{jj'}$) and then summing over $ j'$
we obtain:
\begin{equation}
\rho_{k, e} = \frac{1}{d}\sum
H_{ij}H^*_{ij}|i+q_2,j+q_1\rangle\langle i+q_2,j+q_1|
\end{equation}
Now we use the definition of $H_{ij}:= \frac{1}{\sqrt{d}} \zeta
^{ij}$, to set $H_{ij}H^*_{ij}= \frac{1}{d} $. The last step is
done by a relabelling of the indices $ i+q_2 $ and $ j+q_1 $ to
end with
\begin{equation}
\rho_{k, e} = \frac{1}{d^2}\sum_{l,m} |l,m\rangle\langle l,m| =
\frac{1}{d^2}\openone_{k}\otimes\openone_{e}
\end{equation}

\section{}
In this appendix we investigate the consequences of
replacing the Hadamard gates with an arbitrary unitary gate $U$.
As stated in the text, the Bell state $|\Psi_{0,0}\rangle$ is
invariant under the action of $U\otimes U^*$ for any unitary
operator. Suppose that Alice or Bob use an operator $U$ instead of
$H$, either deliberately or else by unwanted errors in their
gates. To find the information gain of Eve, we need to calculate
as in appendix \ref{apx:a}, the density matrix $ \rho_{k,e} =
\text{tr}_{a,b}(|\Xi_2\rangle \langle \Xi_2|)$, where $|\Xi_2\rangle $ is
\begin{equation}\label{appendix2-a}
|\Xi_2\rangle = \frac{1}{\sqrt{d}} \sum_{i,j,k = 0}^{d-1}
U_{i,j}U_{k,j}^*|i,k,i+q_2, j+q_1\rangle_{a,b,q,e}
\end{equation}
The calculations are similar to appendix \ref{apx:a}, and the final result
is:
\begin{equation}\label{appendix2-b}
\rho_{k,e} = \frac{1}{d} \sum_{i,j = 0}^{d-1}
|U_{i,j}|^2|i+q_2,j+q_1\rangle_{k,e}\langle i+q_2,j+q_1|
\end{equation}
Thus if the operator $U$ shares only the property with the
Hadamard gate that $ |U_{i,j}|^2 = \frac{1}{d}$, then again we
will have $ \rho_{k,e} = \frac{1}{d^2}\openone_{k}\otimes\openone_{e} $ and the
information gain of Eve reduces to zero. In this sense the
protocol is somehow robust against a large number of errors in the
Hadamard gates.

Secondly we can repeat the calculation that lead to the final
density matrix of the key dits in the hands of Bob, eq.
(\ref{final2}), to determine the new QBER. This time we have
\begin{eqnarray}\label{final3}
|\Xi_4\rangle = \sum_{i,k=0}^{d-1}&&
U_{i,i+q_2-q_1+q}U^*_{k,i+q_2-q_1+q}\nonumber\\
&&|i,k,i+q_2-k,i+q+q_2\rangle
\end{eqnarray}
It is again a simple computation to find the density matrix
corresponding to the Key space from this
state:
\begin{eqnarray}\label{final4}
\rho_{k} &:=& \text{tr}_{a,b,e}|\Xi_4\rangle \langle \Xi_4|\nonumber\\
&=&
\frac{1}{d}\sum_{i,k}|U_{i,i+q_2-q_1+q}|^2|U_{k,i+q_2-q_1+q}|^2\nonumber\\
&&|i+q_2-k\rangle_{k}\langle i+q_2-k|
\end{eqnarray}
Again if for $ |U_{i,j}|^2 = \frac{1}{d}$, we will obtain a
completely mixed matrix $ \frac{1}{d}\openone_{k}$, and the same QBER as
with the Hadamard gates.

\bibliography{qkdrev}
\end{document}